\newtheorem{Theorem}{Theorem}
\documentclass[letterpaper, 10 pt, conference]{ieeeconf}
\usepackage{times,amsmath,amssymb,dsfont,graphicx,float}
\IEEEoverridecommandlockouts
\title{\large \bf Submodularity and Optimality of Fusion Rules in Balanced Binary Relay Trees}

\author{Zhenliang~Zhang,
        Edwin~K.~P.~Chong,
        Ali~Pezeshki,
        William~Moran,
        and Stephen~D.~Howard
\thanks{This work was supported in part by AFOSR under Contract FA9550-09-1-0518, and by NSF under Grants ECCS-0700559, CCF-0916314, and CCF-1018472.}
\thanks{Z. Zhang, E. K. P. Chong, and A. Pezeshki are with the Department
of Electrical and Computer Engineering, Colorado State University, Fort Collins, CO 80523-1373, USA
{\tt\small {zhenliang.zhang, edwin.chong, ali.pezeshki@colostate.edu}}
}
\thanks{W. Moran is with the Department of Electrical and Electronic Engineering, The University of Melbourne, Melbourne, VIC 3010, Australia {\tt\small {wmoran@unimelb.edu.au}}
}
\thanks{S. D. Howard is with the Defence Science and Technology Organisation, P.O.
Box 1500, Edinburgh, South Australia 5111, Australia {\tt\small {sdhoward@unimelb.edu.au}}
}
}

\begin{document}

\maketitle
\thispagestyle{empty}
\pagestyle{empty}

\begin{abstract}
We study the distributed detection problem in a balanced binary relay tree, where the leaves of the tree are sensors generating binary messages. The root of the tree is a fusion center that makes the overall decision. Every other node in the tree is a fusion node that fuses two binary messages from its child nodes into a new binary message and sends it to the parent node at the next level. We assume that the fusion nodes at the same level use the same fusion rule. We call a string of fusion rules used at different levels a fusion strategy.
We consider the problem of finding a fusion strategy that maximizes the reduction in the total error probability between the sensors and the fusion center. We formulate this problem as a deterministic dynamic program and express the solution in terms of {Bellman's equations}. We introduce the notion of \emph{string-submodularity} and show that the reduction in the total error probability is a string-submodular function. Consequentially, we show that the greedy strategy, which only maximizes the level-wise reduction in the total error probability, is within a factor $(1-e^{-1})$ of the optimal strategy in terms of reduction in the total error probability.
\end{abstract}

\section{Introduction}
Consider a distributed detection network consisting of a set of sensors and fusion nodes. The objective is to collectively solve a binary hypothesis testing problem. The sensors make observations from a common event, and then communicate quantized messages to other fusion nodes, according to the network architecture. Each fusion node fuses the received messages from its child nodes into a new message and then sends it to the fusion node at the next level for further integration. A final decision is eventually made at a central fusion node, usually called the fusion center. A fundamental question is how to fuse messages at each fusion node such that the fusion center makes the best decision, in the sense of optimizing a global objective function. For example, under the Neyman-Pearson criterion, the objective is to minimize the probability of missed detection with an upper bound constraint on the probability of false alarm; under the Bayesian criterion, the objective is to minimize the total error probability.

The {distributed detection} problem has been investigated extensively in the context of different network architectures.
In the well-studied \emph{parallel network}~\cite{Tenney}--\nocite{Chair,Cham,dec,Tsi,Tsi1,Warren,Vis,Poor,Sah,chen1,Liu,chen,Kas,Hao,Mou,Chong}\cite{BOOK} where sensors communicate with the fusion center directly, with the assumption of independent sensor observations conditioned on either hypothesis, the optimal fusion rule under the Bayesian criterion is simply a likelihood-rate test with a threshold given by the ratio of the prior probabilities.

The \emph{tandem network} has been considered in \cite{tmc}--\nocite{Tang,Tum,tandem,athans}\cite{Venu},
in which each fusion node combines the observation from its own sensor with the message it receives from its child node at one level down, and then transmits the combined message to its parent node at the next level up. We call a collection of fusion rules at  all fusion nodes a \emph{fusion strategy}. Specifically, \cite{Tang} considers the problem of finding the optimal fusion strategy in such a tandem network under both Neyman-Pearson and Bayesian criteria.

The \emph{bounded-height tree network} has been considered in \cite{Tang1}--\nocite{Nolte,tree1,tree2,tree3,Pete,Alh,Will}\cite{Lin}, where the leaves are sensors, the root is the fusion center, and every other node is a fusion node that fuses the messages from its child nodes and sends a new message to its parent node.  In general, finding a fusion strategy that minimizes the total error probability at the fusion center in bounded-height trees is computationally intractable even for a network with moderate number of nodes. Therefore, many recent papers focus on the asymptotic decay rate of the total error probability as the number of sensors goes to infinity. In balanced bounded-height trees where all the leaf nodes are at the same distance from the fusion center, a fusion strategy that $\epsilon$-achieves the optimal decay exponent is studied \cite{tree1}, in which all the fusion nodes at the same level use the same likelihood-ratio test as the fusion rule.

\begin{figure}[htbp]
\centering
\includegraphics[width=3.7in]{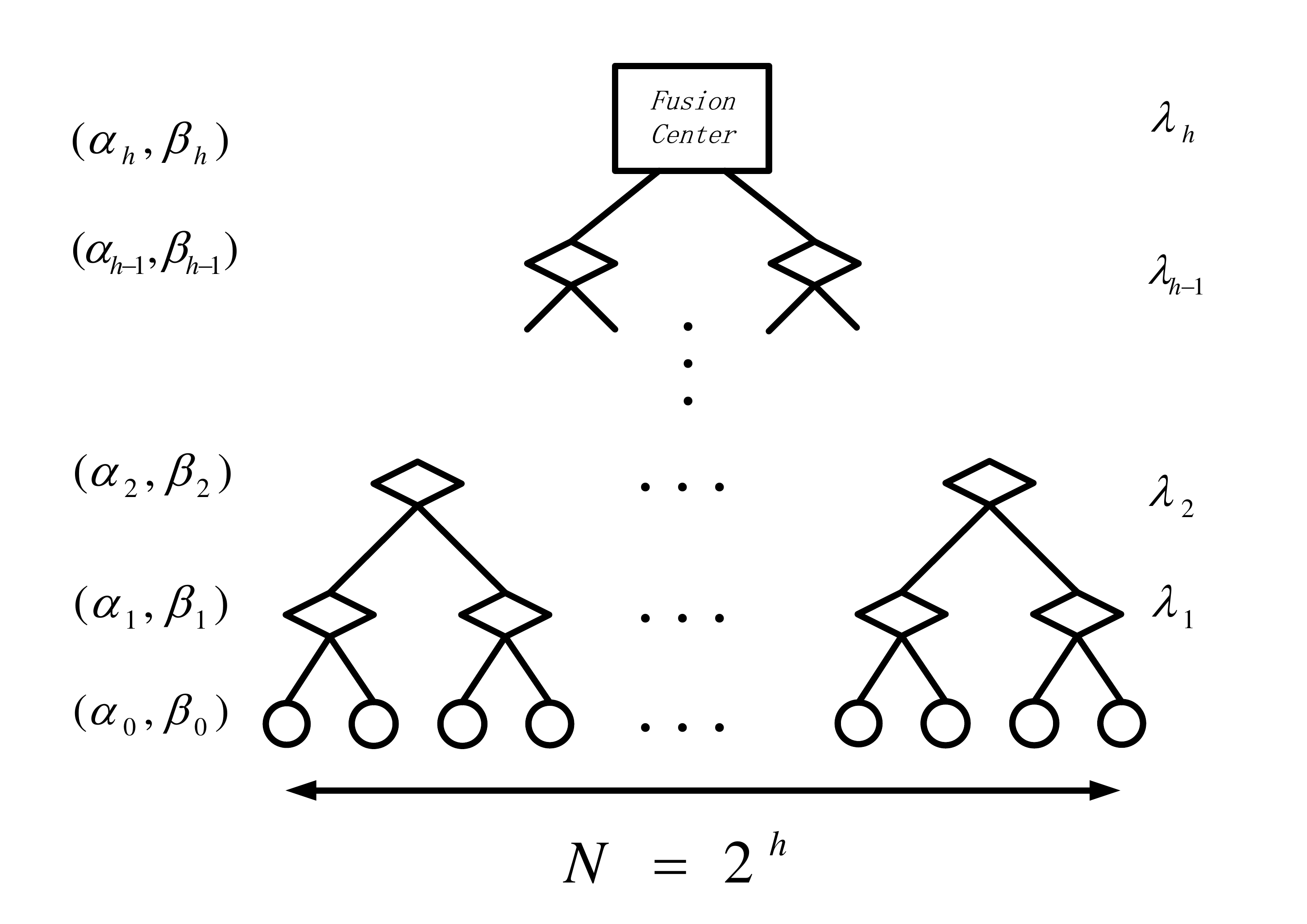}
\caption{A balanced binary relay tree with height $h$. Circles represent sensors making measurements. Diamonds represent relay nodes which fuse binary messages. The rectangle at the root represents the fusion center making an overall decision.}
\label{fig:tree}
\end{figure}

The \emph{unbounded-height tree network} has been considered in \cite{Gubner}--\nocite{Zhang,Zhang2,yash}\cite{Zhang4}. In particular, \cite{Gubner} considered balanced binary relay trees with the structure shown in Fig.~\ref{fig:tree}. In this configuration, the leaf nodes, depicted as circles, are sensors generating binary messages independently and forward these binary messages to their parent nodes. Each node depicted as a diamond is a fusion (relay) node, which fuses the two binary messages received from its child nodes and forwards the new message upward. Ultimately, the fusion center at the root makes an overall decision. This tree is balanced in the sense that all the leaf nodes are at the same distance from the fusion center, and it is binary in the sense that each nonleaf node has two child nodes. This architecture is of interest because it represents the worst-case scenario in the sense that the minimum distance from the sensors to the fusion center is the largest. Assuming that all nonleaf nodes use the same fusion rule, the unit-threshold likelihood-ratio test (ULRT). \cite{Gubner} shows the convergence of detection error probabilities using a Lyaponov method. Under the same assumptions, we further show in \cite{Zhang} that the decay rate of the total error probability is $\sqrt N$, where $N$ is the number of sensors. Under the equally-likely prior probability assumption, ULRT is the locally optimal fusion rule in the sense that  the total error probability of each node is minimized after each fusion. However, we do not expect the strategy consisting of repeated ULRT fusion rules (which we call the \emph{greedy} strategy) to be globally optimal in the sense that the total error probability at the fusion center is minimized.

In this paper we are interested in the following questions:
\begin{itemize} 
\item[1)] What is the globally optimal strategy for balanced binary relay trees?
\item[2)] How much difference in terms of the total error probability is there between the globally optimal strategy and the greedy strategy? 
\end{itemize} 
We answer the first question by formulating the problem as a dynamic program and characterizing the optimal strategy using {Bellman's equations}.
We answer the second question by introducing the notion of string-submodularity and showing that the reduction in the total error probability is a string-submodular function. Subsequently, we show that the reduction in the total error probability achieved by the greedy strategy is at least a factor $(1-e^{-1})$ of that achieved by the globally optimal strategy.

\section{Problem Formulation}

We consider the problem of testing binary hypothesis between $H_0$ and $H_1$ in a balanced binary relay tree, with structure shown in Fig.~\ref{fig:tree}. 
Let $p$ be any fusion node (i.e., $p$ is a nonleaf node). We denote by $C(p)$ the set of child nodes of $p$. Suppose that $p$ receives binary messages $Y_c\in\{0,1\}$ from every $c\in C(p)$ (i.e., from its child nodes), and then summarizes the two received binary messages into a new binary message $Y_p\in\{0,1\}$ using a fusion rule $\lambda_p$:
\[
 Y_p=\lambda^p(\{Y_c: c\in C(p)\}).
\]
 The new message $Y_p$ is then communicated to the parent node (if any) of $p$. Ultimately, the fusion center makes an overall decision.

It turns out that the only meaningful rules to aggregate two binary messages in this case are simply `AND' and `OR' rules defined as follows:
\begin{itemize}
\item AND rule (denoted by $\mathcal A$): a parent node decides $1$ if and only if both its child nodes send~$1$;
\item OR rule (denoted by $\mathcal O$): a parent node decides $0$ if and only if both its child nodes send $0$.
\end{itemize}
Henceforth, we only consider the case where each fusion node in the tree choose a fusion rule from $\mathcal Y:=\{\mathcal A, \mathcal O\}$.

We assume that all sensors are independent with identical Type I error probability $\alpha_0$ and identical Type II error probability $\beta_0$. Moreover, we assume that all the fusion nodes at level $k$ $(k\geq 1)$ use the same fusion rule $\lambda_k$; i.e., for each node $p$ that lies at the $k$th level of the tree, $\lambda^{p}=\lambda_k$.
In this case, all the output binary messages for nodes at level $k$ have the same Type I and Type II error probabilities, which we denote by $\alpha_k$ and $\beta_k$ respectively. Given a fusion rule $\lambda_k$, we can show that the error probabilities evolve as follows:
\begin{equation*}
(\alpha_{k}, \beta_{k}):=\left\{\begin{array}{c}
(1-(1-\alpha_{k-1})^2, \beta_{k-1}^2), \quad \text{if $\lambda_k=\mathcal A$},\\
(\alpha_{k-1}^2, 1-(1-\beta_{k-1})^2), \quad \text{if $\lambda_k=\mathcal O$}.\end{array}\right.
\end{equation*}

\emph{Remark}: Note that the evolution of the error probability pair $(\alpha_k,\beta_k)$ is symmetric with respect to the line $\alpha+\beta=1$. 
Hence, it suffices to consider the case where the initial pair satisfies $\alpha_0+\beta_0<1$. We can derive similar result for the case where $\alpha_0+\beta_0>1$ (e.g., by only flipping the decision at the fusion center).  In the case where $\alpha_0+\beta_0=1$, the Type I and II error probabilities add up to one regardless of the fusion rule used. Hence, this case is not of interest.

Notice that the ULRT fusion rule is either the $\mathcal A$ rule or the $\mathcal O$ rule, depending on the values of the Type I and Type II error probabilities at a particular level of the tree. More precisely, we have
\begin{itemize}
\item If $\beta_k> \alpha_k$, then the ULRT fusion rule is $\mathcal A$;
\item If $\beta_k< \alpha_k$, then the ULRT fusion rule is $\mathcal O$;
\item
If $\beta_k= \alpha_k$, then the total error probability remains unchanged after using $\mathcal A$ or $\mathcal O$. Moreover, the error probability pairs at the next level $(\alpha_{k+1},\beta_{k+1})$ after using $\mathcal A$ or $\mathcal O$ are symmetric about the line $\beta=\alpha$. Therefore, we call both $\mathcal A$ and $\mathcal O$ the ULRT fusion rule in this case.
\end{itemize}

We define a fusion strategy as a string of fusion rules $\lambda_j\in \mathcal Y$ used at levels $j=1,2,\ldots,h$, denoted by $\pi=(\lambda_1,\lambda_{2},\ldots,\lambda_{h})$. Let the collection of all possible fusion strategies with length $h$ be $\mathcal Y^h$:
\[
\mathcal Y^h:=\{\pi=(\lambda_1,\lambda_{2},\ldots,\lambda_{h})| \lambda_j\in \mathcal Y \text{ } \forall j\}.
\]
For a given initial error probability pair $(\alpha_{0},\beta_{0})$ at the sensor level, the pair $(\alpha_{h},\beta_{h})$ at the fusion center (level $h$) is a function of $(\alpha_{0},\beta_{0})$ and the specific fusion strategy $\pi$ used.
We consider the Bayesian criterion in this paper, under which the objective is to minimize the total error probability $\mathbb{P}(H_0)\alpha_h+ \mathbb{P}(H_1)\beta_h$ at the fusion center, where $\mathbb{P}(H_0)$ and $\mathbb{P}(H_1)$ are the prior probabilities of the two hypotheses, respectively. Equivalently, we can find a strategy that maximizes the reduction of the total error probability between the sensors and the fusion center. We call this optimization problem an \emph{$h$-optimal problem}. Without loss of generality, we assume that the prior probabilities are equal; i.e., $\mathbb{P}(H_0)=\mathbb{P}(H_1)=1/2$, in which case the $h$-optimal problem (ignoring a factor of $1/2$) can be written as:
\begin{align}
\begin{array}{l}
\text{maximize }\alpha_0+\beta_0-(\alpha_{h}+\beta_{h})\\
\text{subject to }{\pi\in \mathcal Y^h.}
\end{array}
\label{eq2}
\end{align}
A fusion strategy that maximizes \eqref{eq2} is called the $h$-optimal strategy:
\begin{align*}
{\pi}^o(\alpha_0,\beta_0)  &=\operatorname*{arg\,max}_{\pi\in \mathcal Y^h} (\alpha_0+\beta_0-(\alpha_{h}+\beta_{h}))\\
   &=\operatorname*{arg\,max}_{\pi\in \mathcal Y^h} \sum_{j=0}^{h-1}(\alpha_j+\beta_j-(\alpha_{j+1}+\beta_{j+1})).
\end{align*}


In contrast, the ULRT fusion rule minimizes the step-wise reduction in the total error probability:
\[
\text{ULRT} =\operatorname*{arg\,max}_{\lambda_i \in \mathcal Y} (\alpha_i+\beta_i-(\alpha_{i+1}+\beta_{i+1}))  \quad\forall i.
\]
Because of the equal prior probability assumption, a \emph{maximum a posteriori} (MAP) fusion rule is the same as the ULRT fusion rule. In this context, we call a fusion strategy consisting of repeated ULRT fusion rules a ULRT (greedy) strategy.

In the next section, we derive the $h$-optimal fusion strategy for balanced binary relay trees with height $h$ using Bellman's equations. We then show that the 2-optimal strategy is equivalent to the ULRT strategy. Moreover, we show that the reduction of the total error probability is a string-submodular function (as defined in Section III-C), which implies that the greedy strategy is close to the optimal fusion strategy in terms of the reduction in the total error probability.

\section{Main Results}

\subsection{Dynamic Programming Formulation}
In this section, we formulate the problem of finding the optimal fusion strategy using a deterministic dynamic programming model. First we define the necessary elements of this dynamic model.
\begin{itemize}
\item[I.] \emph{Dynamic System}: 
We define the error probability pair at the $k$th level $(\alpha_k,\beta_k)$ as the system state, denoted by $s_k$. Notice that $\alpha_k$ and $\beta_k$ can only take values in the interval $[0,1]$. Therefore, the set of all possible states is $\{(\alpha,\beta)>0|\alpha+\beta<1\}$.
Moreover, given the fusion rule, the \emph{state transition function} is deterministic. If we choose $\lambda_k=\mathcal A$, then
    \[(\alpha_{k},\beta_{k})=(1-(1-\alpha_{k-1})^2,\beta_{k-1}^2).\]
    On the other hand, if we choose $\lambda_k=\mathcal O$, then
    \[(\beta_{k},\alpha_{k})=(1-(1-\beta_{k-1})^2,\alpha_{k-1}^2).\]
\item[II.] \emph{Rewards}: 
At each level $k$, we define the instantaneous reward to be the reduction of the total error probability after fusing with $\lambda_k$:
    \[r(s_{k-1},\lambda_k)=(\alpha_{k-1}+\beta_{k-1})-(\alpha_{k}+\beta_{k}),\]
     where $\alpha_{k}$ and $\beta_{k}$ are functions of the previous state $s_{k-1}$ and the fusion rule $\lambda_k$.

\end{itemize}

Let $v_{h-k} (s_k)$ be the cumulative reduction of the total error probability if we start the system at state $s_k$ at level $k$ and the strategy $(\lambda_{k+1},\lambda_{k+2}\ldots,\lambda_h)\in \mathcal Y^{h-k}$ is used. Following the above definitions, we have
\[
v_{h-k} (s_k)=\sum_{j=k+1}^h r(s_{j-1},\lambda_j) .
\]
If we let $k=0$, that is, we start calculating the reduction from the sensor level, then the above cumulative reward function is the same as the global objective function defined in Section~II.
Therefore, for given initial state $s_0$, we have to solve the following optimization problem to find the global optimal strategy over horizon $h$:
\[
v^o_{h} (s_0)=\max_{\pi\in \mathcal Y^h}\sum_{j=1}^h r(s_{j-1},\lambda_j) .
\]
The globally optimal strategy ${\pi}^o$ is
\[
{\pi}^o (s_0)=\operatorname*{arg\,max}_{\pi\in \mathcal Y^h}\sum_{j=1}^h r(s_{j-1},\lambda_j) .
\]
Notice that $s_k$ depends on the previous state $s_{k-1}$ and the fusion rule $\lambda_k$. Hence we write the state at level $k$ to be $s_k|_{s_{k-1},\lambda_k}$. The solution of the above optimization problem can be characterized using {Bellmam's equations}, which state that
\[
v_{h}^{o} (s_0)=\max_{\lambda_1\in \mathcal Y}\left[r(s_0,\lambda_1)+v_{h-1}^{o} (s_1|_{s_0,\lambda_1}) \right]
\]
\[
\lambda_1^o (s_0)=\operatorname*{arg\,max}_{\lambda_1\in \mathcal Y}\left[r(s_0,\lambda_1)+v_{h-1}^{o} (s_1|_{s_0,\lambda_1}) \right],
\]
where $\lambda_1^o(s_0)$ is the first element of the optimal strategy $\pi^o$. Recursively, the solution of the optimization problem is
\[
v_{h-(k-1)}^{o} (s_{k-1})=\max_{\lambda_{k}\in \mathcal Y}\left[r(s_{k-1},\lambda_{k})+v_{h-k}^{o} (s_{k}|_{s_{k-1},\lambda_{k}}) \right].
\]
Moreover, the $k$th element of the optimal strategy $\pi^o$ is
\begin{align*}
\lambda_{k}^o (s_{k-1})=
\operatorname*{arg\,max}_{\lambda_{k}\in \mathcal Y}\left[r(s_{k-1},\lambda_{k})+v_{h-k}^{o} (s_{k}|_{s_{k-1},\lambda_{k}}) \right].
\end{align*}

\emph{Remark}: The above formulation can easily be generalized to the node and  link failure case  \cite{Zhang2} and even more complicated architectures (e.g., bounded-height trees \cite{tree1} and $M$-ary relay trees \cite{Zhang4}) simply by changing the state transition functions and the set of all possible fusion rules. Also, we can generalize the formulation to non-equal prior probability scenarios.

The complexity of the explicit solution to Bellman's equations grows exponentially with respect to the horizon $h$. Therefore, it is usually intractable to compute the $h$-optimal strategy if $h$ is sufficiently large. An alternative strategy is the ULRT strategy, which consists of repeating ULRT fusion rule at all levels. We have shown in~\cite{Zhang} that the decay rate of the total error probability with this strategy is $\sqrt N$. Next we study whether the ULRT strategy is the same as the $h$-optimal strategy. If not, does the ULRT strategy provide a reasonable approximation of the $h$-optimal strategy?

\subsection{2-optimal Strategy}
In this section, we show that the 2-optimal strategy is the same as the ULRT strategy. Moreover, we give an counterexample which shows that the ULRT strategy is not 3-optimal.

Consider the 2-optimal problem in the balanced binary relay tree with height 2:
\begin{align*}
v^o_{2} (s_0)&=\max_{\pi\in \mathcal Y^2}\sum_{j=1}^2 r(s_{j-1},\lambda_j),
\end{align*}
where $\mathcal Y^2=\{(\mathcal A,\mathcal A), (\mathcal A,\mathcal O), (\mathcal O,\mathcal O), (\mathcal O,\mathcal A)\}$. The 2-optimal strategy in this case is
 \begin{align*}
{\pi}^o (s_0)&=\operatorname*{arg\,max}_{\pi\in \mathcal Y^2}\sum_{j=1}^2 r(s_{j-1},\lambda_j).
\end{align*}
 We have the following theorem. (Because of lack of space, many of the proofs here are omitted.)

\begin{Theorem}
A strategy ${\pi}$ is 2-optimal if and only if ${\pi}$ is the ULRT strategy.
\end{Theorem}

This result also applies to any sub-tree with height 2 within a balanced binary relay tree with arbitrary height $h>2$. However, the ULRT strategy is not in general optimal for multiple levels; i.e., $h>2$, as the following counter-example for $h=3$ shows.

\begin{figure}[htbp]
\centering
\includegraphics[width=3.6in]{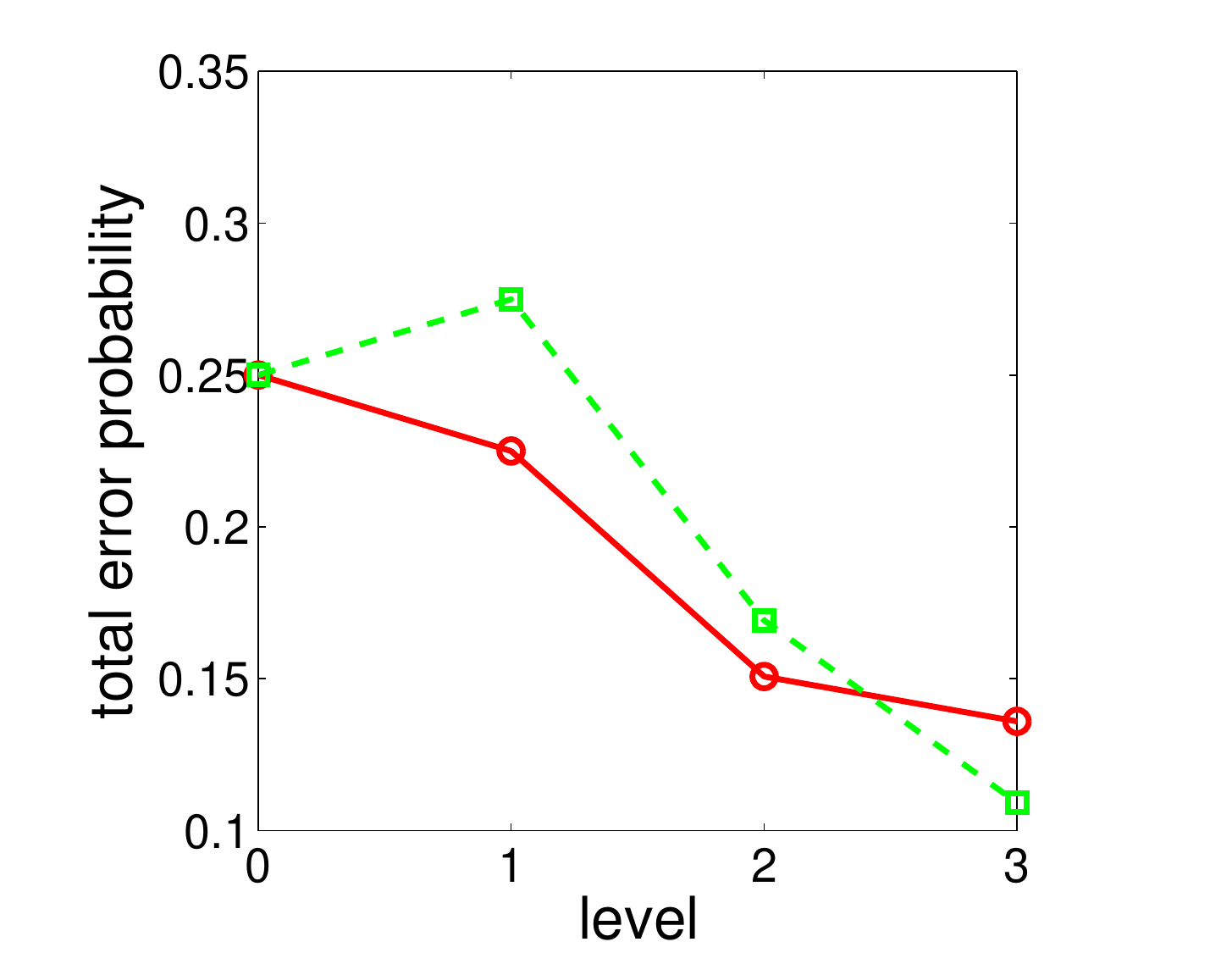}
\caption{Comparison of the ULRT strategy and the 3-optimal strategy. The solid (red) line represents the error probability curve using the ULRT strategy. The dashed (green) line represents the error probability curve using the 3-optimal strategy.}
\label{fig:ex}
\end{figure}

Let the initial state be $(\alpha_0,\beta_0)=(0.2,0.3)$, in which case the ULRT strategy is $(\mathcal A,\mathcal O, \mathcal A)$. As shown in Fig.~\ref{fig:ex}, the solid (red) line denotes the total error probabilities at each level up to 3. However, the 3-optimal strategy in this case is $(\mathcal O,\mathcal A, \mathcal A)$. The total error probability curve of this strategy is shown as a dashed (green) line in Fig.~\ref{fig:ex}. Similar counterexamples can be found for cases where $h>3$. Hence, the ULRT strategy is not in general $h$-optimal for $h\geq 3$. In the next section, we will introduce and employ the notion of string-submodularity to quantify the gap in performance between optimal and ULRT strategies for $h\geq 3$.

\subsection{String-submodularity}

Submodularity of functions over finite sets plays an important role in combinatorial optimization. It has been shown that the greedy strategy provides at least a constant-factor approximation to the optimal strategy. For example, the celebrated result of Nemhauser \emph{et al.}~\cite{Nem} states that for maximizing a monotone submodular function $F$ over a uniform matroid such that $F(\emptyset)=0$ (here $\emptyset$ denotes the empty set), the value of the greedy strategy is no less than a factor $(1-e^{-1})$ of that of the optimal strategy.

Note that the submodular functions studied in most previous papers are defined on the power set of a given finite set. However, many stochastic optimization problems concern optimizing objective functions over a finite horizon, where we have to choose an action from a given finite set at each iteration. In these cases, the objective function usually depends on the order of actions, and repeating the same action is allowed. Hence, we cannot  directly apply the result of Nemhauser \emph{et al.} \cite{Nem}.

For objective functions defined on strings (finite-length sequences), \cite{streeter} and \cite{Saeed} provide some sufficient conditions such that the greedy strategy achieves a good approximation for maximizing the string function. In this paper, we improve these results by providing sufficient conditions that are weaker than those in \cite{streeter} and \cite{Saeed}.

Next we formulate the maximization problem using submodular functions defined on strings.

\begin{itemize}
\item[I.] 
\emph{String}:
Consider a finite set $\mathbb{A}$ of possible actions. For each step, we choose an action from $\mathbb{A}$. Let $A=(a_1,a_2,\ldots,a_k)$ be a string of actions taken over $k$
steps, where $a_i\in \mathbb{A}$ for all $i$. Let the set of all strings of actions be
\[
\mathbb A^*=\{(a_1,a_2,\ldots,a_k)|k=0,1,\ldots \text{ and }  a_i\in \mathbb A \text{ }\forall i\}.
\]
Note that $k=0$ corresponds to the empty string (no action taken), denoted by $\emptyset$.

\item[II.]
\emph{String length}:
For a given string $A=(a_1,a_2,\ldots,a_k)$, we define its \emph{string length} as $k$, denoted $|A|=k$.

\item[III.]
\emph{String concatenation}:
Let $M=(a_1^m,a_2^m,\ldots, a_{k_1}^m)$ and $N=(a_1^n,a_2^n,\ldots, a_{k_2}^n)$ be two strings in $\mathbb A^*$. 
We define concatenation as follows:
\[
M\oplus N= (a_1^m,a_2^m,\ldots, a_{k_1}^m,a_1^n,a_2^n,\ldots, a_{k_2}^n).
\]

\item[IV.]
\emph{String dominance}:
Let $M$ and $N$ be two strings in $\mathbb A^*$. We write $M\preceq N$ if we have
\[
N=M\oplus(a_1,a_2,\ldots, a_j),
\]
where $j\in \{0,1,\ldots\}$ and $a_i\in \mathbb A$ for all $i$. In other words, $M$ is a \emph{prefix} of $N$.

\item[V.]
\emph{String-submodularity}:
A function from strings to real numbers, $f: \mathbb A^*\to \mathbb R$, is string-submodular if
\begin{itemize}
\item[i.] $f$ has the \emph{monotone} property; i.e.,
\[f(M)\leq f(N), \quad \forall M\preceq N \in \mathbb A^*.
\]
\item[ii.] $f$ has \emph{diminishing-return} property; i.e.,
\begin{align*}
\nonumber
&f(M\oplus (a))-f(M) \geq f(N\oplus (a))-f(N), \\
&\forall M\preceq N \in \mathbb A^*, \forall a\in \mathbb A.
\end{align*}
\end{itemize}
Note that the diminishing-return property here only requires concatenating one more action. 

\item[VI.]  
\emph{Globally optimal solution}:
Consider the problem of finding a string that maximizes $f$ under the constraint that the string length is not larger than $K$. Because the function $f$ is monotone, it suffices to consider the stronger constraint with fixed length $K$:
\begin{align}
\begin{array}{l}
\text{maximize  }   f(M) \\
\text{subject to  } M\in\mathbb A^*, |M|=K.
\end{array}
\label{eqn:1}
\end{align}
\item[VII.]
\emph{Greedy solution}:
A string $G=(a_1^*,a_2^*,\ldots,a_{|G|}^*)$ is called \emph{greedy} if
\begin{align*}
a_i^*=&\operatorname*{arg\,max}_{a_i\in \mathbb A} f((a_1^*,a_2^*,\ldots,a_{i-1}^*,a_i))\\
&-f((a_1^*,a_2^*,\ldots,a_{i-1}^*)) \quad \forall i=1,2,\ldots,|G|.
\end{align*}

\end{itemize}

For a deterministic dynamic system, the diminishing return property can be simplified as follows.

\emph{Lemma 1}: For any $M, \text{ }N\in \mathbb A^*$ and $a\in \mathbb A$, we have
\begin{align*}
\nonumber
&f(M\oplus (a))-f(M) \geq f(N\oplus (a))-f(N)
\end{align*}
if and only if
\[
f((a_0) \oplus (a))-f((a_0)) \geq f((a))-f(\emptyset) \quad \forall a_0.
\]


\begin{Theorem} Consider a submodular function $f: \mathbb A^*\to\mathbb R$ such that
\begin{itemize}
\item[i.] $f(\emptyset)=0$;
\item[ii.] For any greedy strings $G$ with a length less than $K$ and optimal strings $O$ (optimal with respect to \eqref{eqn:1}),
$f(G\oplus O) \geq f(O)$.
\end{itemize}
Then any greedy string $G_K$ of length $K$ satisfies
\[
f(G_K)> (1-e^{-1}) f(O).
\]
\end{Theorem}
%

\subsection{Application to Distributed Detection}

We consider balanced binary relay trees with even heights. Again we assume that the nodes at the same level use the same fusion rule. Moreover, we assume that two fusion rules $\Lambda$ of consecutive levels are chosen from the following set $\mathcal Z =\{(\mathcal A, \mathcal O) , (\mathcal O, \mathcal A)\}$. Let $\Pi=(\Lambda_1,\Lambda_2, \ldots, \Lambda_h)$ be a fusion strategy, where $\Lambda_i\in \mathcal Z$ for all $i$. Let $\mathcal Z^*$ be the set of all possible strategies (strings); i.e., $\mathcal Z^*=\{(\Lambda_1,\Lambda_2, \ldots, \Lambda_h)| h=0,1,\ldots \text{ and }  \Lambda_i\in \mathcal Z \text{ }\forall i\}$. Here we only prove the case where the prior probabilities are equally likely. The following analysis easily generalizes to non-equal prior probabilities. Given the two types of error probability $(\alpha_0,\beta_0)$ at level 0, the reduction of the total error probability after applying a strategy $\Pi$ is
\[
u(\Pi)=\alpha_0+\beta_0-(\alpha_{2h}(\Pi)+\beta_{2h}(\Pi)),
\]
where $\alpha_{2h}$ and $\beta_{2h}$ represent the Type I and II error probabilities after fusion using $\Pi$.

Next we show that $u$ is a string-submodular function.

\emph{Proposition 2}:  The function $u$: $\mathcal Z^*\rightarrow \mathbb{R}$ is string-submodular.

For a balanced binary relay trees with height $2K$,
the global optimization problem is to find a strategy $\Pi \in\mathcal Z^*$ with length $K$ such that the above reduction is maximized; that is
\begin{align}
\begin{array}{l}
\text{maximize  }   u(\Pi) \\
\text{subject to  } \Pi\in\mathcal Z^*, |\Pi|=K.
\end{array}
\label{eq3}
\end{align}

We have shown that the reduction of the total error probability $u$ is a string-submodular function. Moreover, we know that the total error probability does not change if there is no fusion; i.e.,
\[
u(\emptyset)=0.
\]
Therefore, we can employ Theorem 2 to the above maximization problem \eqref{eq3}.

Consider a balanced binary relay tree with height $2K$. We denote by $u(G_K)$ the reduction of the total error probability after using the greedy strategy. We have shown that the ULRT strategy is 2-optimal. We have also shown in \cite{Zhang} that the ULRT strategy only allows at most two identical consecutive fusion rules. Hence, we can conclude that a strategy is the ULRT strategy if and only if it is the greedy strategy. We denote by $u(O)$ the reduction of the total error probability using the optimal strategy. We have the following theorem.

\begin{Theorem}
Consider a balanced binary relay tree with height $2K$. We have
\[
(1-e^{-1})u(O)< u(G_K) \leq u(O).
\]
\end{Theorem}

%
%
%
%
%
%
%
%

\emph{Remark}: Recall that the fusion strategy is a string of fusion rules chosen from $\mathcal Z=\{(\mathcal A, \mathcal O), (\mathcal O, \mathcal A)\}$. Thus, the strategies we considered in this section have at most two consecutive repeated fusion rules. For example, the strategy $(\mathcal A, \mathcal A, \mathcal A, \ldots)$ is not considered. It is easy to show that with repeating identical fusion rule, the total error probability goes to 1/2. Therefore, it is  reasonable to rule out this situation.

%

\section{Concluding Remarks}

We study the problem of finding a fusion strategy that maximizes the reduction of the total error probability in balanced binary relay trees. We formulate this optimization problem using deterministic dynamic programming and characterize its solution using Bellman's equations. Moreover, we show that the reduction of the total error probability is a string-submodular function. Therefore, the ULRT strategy, which is a string of repeated ULRT fusion rules, is close to the globally optimal strategy in terms of the reduction in the total error probability.

Future work includes studying the overall optimal strategy for other architectures (e.g., bounded-height trees \cite{tree1} and $M$-ary relay trees). We would like to derive the optimal strategy in balanced binary relay trees with node and link failures. In this paper, we assume that all the nodes at the same level use identical fusion rule. What about the case where the nodes at each level are allowed to use different fusion rules? Moreover, what can we say about trees with correlated sensor measurements? These questions are currently been investigated.

\bibliographystyle{IEEEbib}

\end{document}